# Monitoring Nitric Oxide in Trigeminal Neuralgia Rats with a Cerium Single-Atom Nanozyme Electrochemical Biosensor

*Kangling Tian,*[†][a] *Fuhua Li,*[†][c] *Ran Chen,*[†][a] *Shihong Chen,*[d] *Wenbin Wei,*[e] *Yihang Shen,*[f] *Muzi Xu,*[b] *Chunxian Guo,*[a] *Luigi G. Occhipinti,\**[a,b] *Hong Bin Yang,\**[a] *Fangxin Hu\**[a,b]

K. Tian, R Chen, Prof. C.X. Guo, Prof. H.B. Yang, Prof. F.X. Hu

School of Materials Science and Engineering, Suzhou University of Science and Technology, Suzhou 215009, China

E-mail: yanghb@mail.usts.edu.cn, hufx278@usts.edu.cn

M.Z. Xu, Prof. L.G. Occhipinti, Prof. F.X. Hu

Electrical Engineering Division, Department of Engineering, University of Cambridge, Cambridge, UK

E-mail: lgo23@cam.ac.uk

Dr. F.H. Li

School of Chemistry, Southwest Jiaotong University, Chengdu 610031, China

Prof. S.H. Chen

School of Chemistry and Chemical Engineering, Southwest University, Chongqing, 400715, China

W.B. Wei

Department of Oral Surgery, Shanghai Ninth People's Hospital, School of Medicine; Shanghai Jiao Tong University; National Center for Stomatology; National Clinical Research Center for Oral Diseases; Shanghai Key Laboratory of Stomatology; Shanghai Research Institute of Stomatology, Shanghai, China

Dr. Y.H. Shen

Suzhou Ninth People's Hospital, Suzhou, 215200, China

[†]Contributed equally to this work.

Keywords: Trigeminal neuralgia, cerium single-atom nanozyme, nitric oxide, real-time assessment

**Abstract:** Trigeminal neuralgia (TN) is the most common neuropathic disorder; however, its pathogenesis remains unclear. A prevailing theory suggests that nitric oxide (NO) may induce nerve compression and irritation via vascular dilation, thereby being responsible for the condition, making real-time detection of generated NO critical. However, traditional evaluations of NO rely on indirect colorimetric or chemiluminescence techniques, which offer limited sensitivity and spatial resolution for its real-time assessment in biological environments. Herein, we reported the development of a highly sensitive NO electrochemical biosensor based cerium single-atom nanozyme ($Ce_1$-CN) with ultrawide linear range from 1.08 nM to 143.9 μM, and ultralow detection limit of 0.36 nM, which enables efficient and real-time evaluation of NO in TN rats. In-situ attenuated total reflection surface-enhanced infrared spectroscopy combined with density functional theory calculations revealed the high-performance biosensing mechanism, whereby the Ce centers in $Ce_1$-CN nanoenzymes adsorb NO and subsequently react with $OH^-$ to form $*HNO_2$. Results demonstrated that NO concentration was associated with TN onset. Following carbamazepine treatment, NO production from nerves decreased, accompanied by an alleviation of pain. These findings indicate that the biosensor serves as a valuable tool for investigating the pathogenesis of TN and guiding subsequent therapeutic strategies.

# 1. Introduction

Trigeminal neuralgia (TN) is a chronic and debilitating neuropathy characterized by sudden, brief, intense shooting, stabbing, or shock-like pain in the face[1-3]. Typical TN is believed to be caused by vascular compression of the trigeminal nerve, which can cause damage and demyelination, leading to hyperexcitability and the generation of ectopic action potentials[4-7]. Although the molecular mechanism of TN's pathophysiology remains unclear, the dominant theory suggests that nitric oxide (NO), acting as a vasodilator factor, may cause vasodilation of vascular smooth muscle, which can change the mechanical environment around the trigeminal nerve, resulting in nerve compression or stimulation[8-13]. Therefore, the evaluation of NO generated by the trigeminal nerve is of great significance for elucidating the pathophysiology mechanism of TN[14-15]. However, due to its high reactivity, short half-life (3-5 seconds), and tendency to react with thiols, oxygen, free metal ions and heme-containing proteins[16], real-time quantitative detection of NO remains a significant technical challenge. Various methods such as chemiluminescence[17], fluorescence[18], UV-Vis spectroscopy[19], electrochemical biosensor[20], and electron spin resonance[21] have been developed to break the bottleneck of quantitative detection of NO. However, most of the techniques require complex instruments and are ill-suited for real time analysis in complex media. Electrochemical sensing has attracted much attention due to its low cost, fast response speed, strong miniaturization ability and excellent sensitivity, which is expected to construct an accurate NO sensor with high sensitivity for real time monitoring.

Single-atom nanozymes (SAEs) have become a cutting-edge research topic owing to the adjustable electronic structure of their central metal atoms, which can regulate the adsorption energy of reactants and reaction intermediates. SAEs with different transition metal centers (such as Fe, Co, Ni, Mn and Cu) have attracted much attention due to their excellent biocompatibility and efficient biomimetic catalysis[22-24]. Zhou et al.[25] report on an electrochemical sensor based on Ni-SAE for the detection of NO in a living cell environment. Hu et al.[26] developed an integrated nanoelectronic system with Co-SAE to detect NO produced by various organs of mice. However, there are still some shortcomings in detecting NO produced by complex biological systems, such as stability, sensitivity and selectivity of the sensing materials are still need further improved. The rare earth element cerium (Ce), characterized by its unique 4f shell structure, possesses multiple transition states with significant redox properties. Moreover, the reversible redox cycling between $Ce^{3+}$ and $Ce^{4+}$ can promote the electrochemical oxidation of NO[27-28]. Thus, Ce-based SAEs exhibited great potential for real time NO monitoring.

In this work, an electrochemical sensor based on cerium single atom materials (Ce$_1$-CN) were developed for the detection of NO, which exhibiting an ultrawide linear range detection range with a low detection limit. The underlying reaction mechanism was systematically investigated by in-situ attenuated total reflection surface-enhanced infrared spectroscopy (ATR-SEIRAS) combined with density functional theory (DFT) calculations. The results show that the Ce$_1$-CN nanozyme selectively adsorbs NO molecules through surface active sites, and the adsorbed NO reacts with water to promote the oxidation of NO, leading to the formation of nitrite. Given the unique performance of Ce$_1$-CN material for NO, the developed biosensor successfully enabled real time and accurate monitoring of NO concentrations generated by the trigeminal nerve of normal rats, TN rats, and drug treated TN rats (**Figure 1**). The result showed that when rats suffered from TN, greater amounts of NO were generated in the three regions of the trigeminal nerve. After treatment with carbamazepine, the NO production in these three regions decreased, accompanied by a significant alleviation of pain symptoms in the rats.

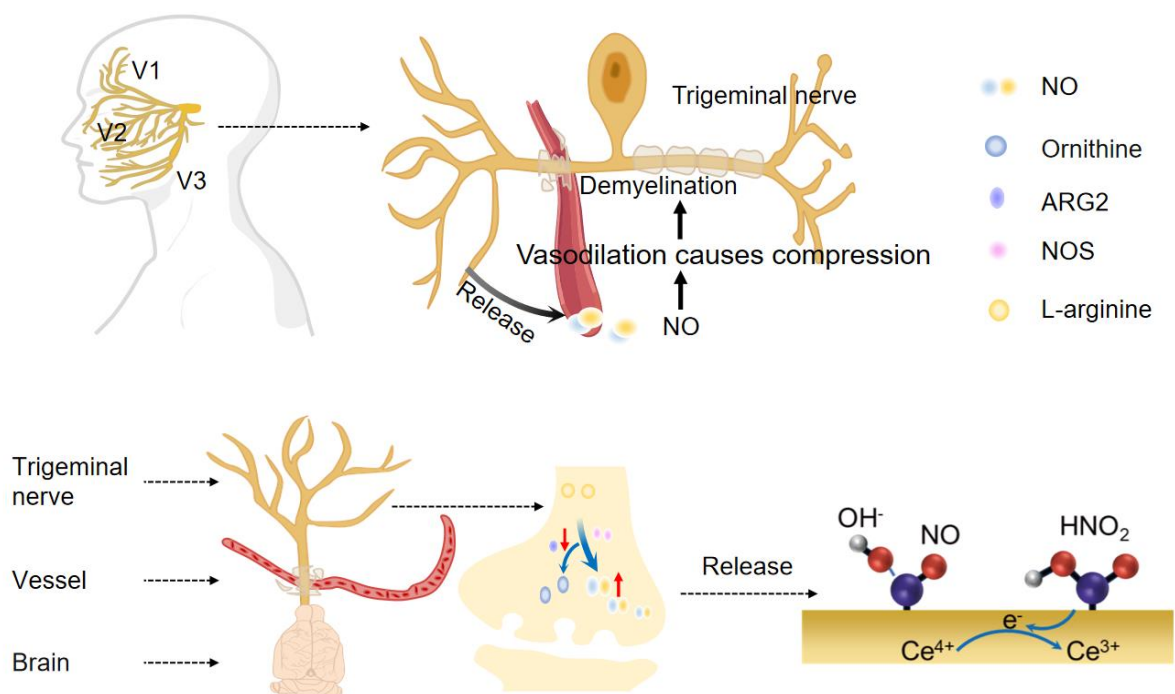

**Figure 1.** Schematic diagram of the relationship between TN and NO, and detection principle of Ce$_1$-CN biosensor.

## 2. Results and Discussion

### 2.1. Structural characterization of Ce$_1$-CN

The Ce$_1$-CN nanozyme was synthesized through pyrolysis of C$_3$H$_6$N$_6$ and Ce(NO$_3$)$_3$·6H$_2$O as precursor (Figure S1, Supporting Information). The morphology of Ce$_1$-CN was characterized by SEM and TEM. SEM (Figure S2, Supporting Information) image shows that $Ce_1$-CN behaves a lamellar smooth structure. And the TEM image (**Figure 2a**) highlights the layered structure of the $Ce_1$-CN. The EDS images show that C, N and Ce are evenly distributed in the sample (Figure 2b). Furthermore, HAADF-STEM was applied to analyze the atomic dispersion of Ce atoms over the $Ce_1$-CN (Figure 2c). Definite bright spots could be observed representing the Ce evenly atomically dispersed on $C_3N_4$. XRD was further used to investigate the crystal structure of the $Ce_1$-CN, and whether Ce ions aggregate during pyrolysis. As shown in Figure 2d, for both $C_3N_4$ and $Ce_1$-CN, diffraction peaks at 12.9° and 27.7° existed corresponding to the (002) and (001) planes of $C_3N_4$, respectively, which consistent with the standard card (JCPDSNO. 87-1526) [29]. Results suggested that there is no obvious metal Ce diffraction peak, which proves that there are no metal Ce atoms and co-derived particles in $Ce_1$-CN, showing atomic-level dispersion of Ce in the sample.

The surface chemical compositions and states of $Ce_1$-CN were studied by XPS (Figure 2e-g and S3, Supporting Information). The Ce 3d narrow spectrum (Figure 2e) could be deconvoluted into the Ce ions in different valence states, the peaks at 904.7 eV and 886.1 eV are assigned to the main $3d_{3/2}$ and $3d_{5/2}$ peaks of Ce (IV), respectively[30]. While peaks at 900.3 eV and 882.2 eV are likely to be composed of the $3d_{3/2}$ and $3d_{5/2}$ of Ce (III) spin-orbit splitting[31]. The peak of $3d_{3/2}$ in Ce-CN is slightly shifted to higher binding energy compared to $CeCl_3$ (III), which indicates that the electronic interaction between Ce and N has changed the electronic structure of Ce to higher valence. The spectrum of N1s (Figure 2f) of $Ce_1$-CN and $C_3N_4$ could be deconvoluted into four fitted peaks, which would be assigned to the pyridine N and pyrrolic N (398.6 eV and 399.0 eV, 398.5 eV and 398.8 eV), graphite N (400.2 eV, 400.2 eV), and oxidized N (401.2 eV, 401.2 eV), respectively[32]. Results show that the pyridine N and pyrrolic N peaks in $Ce_1$-CN shift to higher binding energies by 0.1 and 0.2 eV compared with $C_3N_4$ due to chemical bonding-induced electron transfer, implying a Ce-pyridine N bonding in $Ce_1$-CN [33]. The nearly identical C1 spectra of $Ce_1$-CN and $C_3N_4$ (Figure 2g) indicated that the Ce atoms did not form direct bonds with the C atoms in the $C_3N_4$ support.

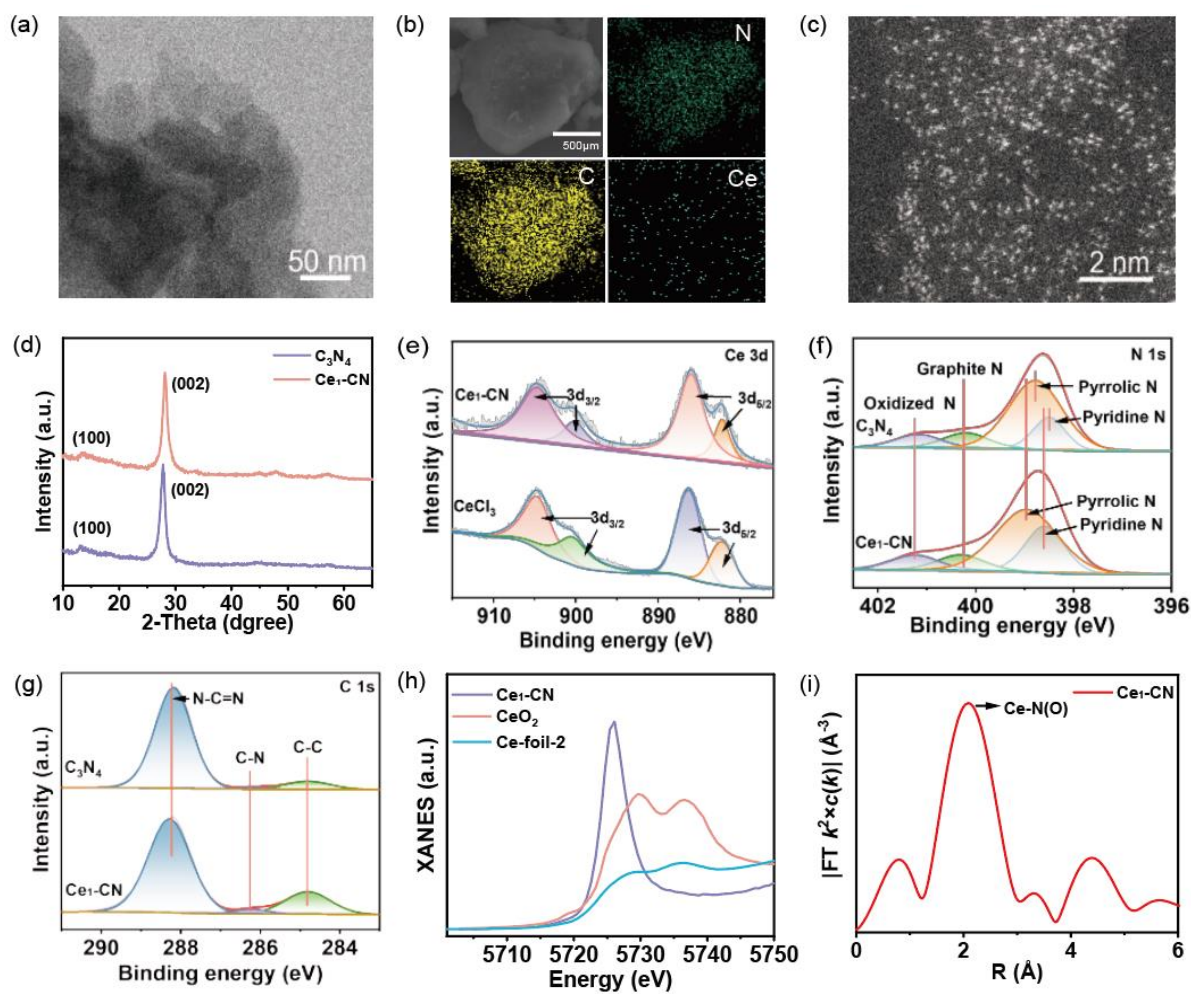

**Figure 2.** (a) TEM (b) EDS and (c) HAADF-STEM images of $Ce_1$-CN. (d) XRD pattern for $Ce_1$-CN and $C_3N_4$. (e) High-resolution XPS of Ce 3d in $Ce_1$-CN and $CeCl_3$. High-resolution XPS of $Ce_1$-CN and $C_3N_4$ (f) N 1s, and (g) C 1s spectra. (h) Ce L3-edge XANES spectra of $Ce_1$-CN, $CeO_2$ and Ce-foil-2. (i) FT-EXAFS spectra of $Ce_1$-CN.

The X-ray absorption near-edge structure (XANES) spectrum and Fourier transformed extended X-ray absorption fine structure (EXAFS) spectrum for L3-edge of $Ce_1$-CN were recorded to further determine its local electronic and coordination structures. The XANES spectrum (Figure 2h) at the Ce L3-edge of $Ce_1$-CN exhibits a sharp single peak at 5726.1 eV, which is different from that of $CeO_2$ (reference sample) with two peaks at 5729.6 and 5736.6 eV[34]. Combined with Ce 3d XPS result, a mixed valence state of $Ce^{3+}$ and $Ce^{4+}$ exists for Ce atom in $Ce_1$-CN can be confirmed. The FT-EXAFS spectra of $Ce_1$-CN is shown in Figure 2i, a prominent Ce-N(O) characteristic peak emerges at approximately 2.08 Å, reflecting the coordination interaction between Ce atoms and N atoms. Furthermore, the absence of peak at 3.68 Å associated with the Ce-Ce coordination indicated the Ce atoms atomically dispersed on the $C_3N_4$[35], in consistent with the HAADF-STEM result.

## 2.2. Electrocatalytic behavior of Ce$_1$-CN on NO

Cyclic voltammetry (CV) and differential pulse voltammetry (DPV) techniques were used to evaluate the sensor performance of Ce$_1$-CN nanozymes. As shown in **Figure 3a**, the CV results showed that Ce$_1$-CN/GCE (light purple line) and C$_3$N$_4$/GCE (orange line) have no obvious redox reactions in 0.01 M PBS. After 72 μM NO was added in 0.01 M PBS, Ce$_1$-CN/GCE exhibited the much more negative peak potential of 0.95 V compared with C$_3$N$_4$ (1.11 V), suggesting that Ce$_1$-CN has the better electrocatalytic activity for the oxidation of NO. In addition, Ce$_1$-CN/GCE (blue line) exhibited a much higher current response value toward NO oxidation compared with C$_3$N$_4$ (purple line), indicated that it was more effective in catalysis of NO. Besides, the DPV (Figure S4, Supporting Information) plot also showed the same results as that of CV. The electrochemical response of the modified electrodes was further investigated using linear sweep voltammetry (LSV) (Figure S5, Supporting Information). The results demonstrate that the onset potential for NO oxidation at Ce$_1$-CN/GCE (green curve, 0.69 V) exhibits a significant 70 mV negative shift compared to that at C$_3$N$_4$/GCE (orange curve, 0.76 V), confirming that Ce$_1$-CN possesses enhanced catalytic activity. The feasibility of Ce$_1$-CN for the detection of NO was further evaluated using CV (Figure S6, Supporting Information). After the addition of 0 μM, 36 μM, 72 μM, 108 μM, 144 μM, and 180 μM NO, the response current values of Ce$_1$-CN/GCE increased gradually, indicating good feasible and stability. In order to further investigate the kinetic process of NO oxidation catalyzed by Ce$_1$-CN, CV curve scanning was performed in the sweep speed range of 10~1000 mV s$^{-1}$. As shown in Figure 3b, the oxidation peak current increased with the increase of the scan rate, and there was a linear relationship between the anodic peak current and the scan rate, I (μA) =0.83 + 2.61224 S (mV s$^{-1}$) (R$^2$=0.99, Figure S7, Supporting Information), which indicated the existence of a surface adsorption relationship for the whole kinetic process.

The experimental conditions were further optimized for the constructed Ce$_1$-CN sensor toward NO sensing. The effect of pH on the electrocatalytic oxidation performance of Ce$_1$-CN nanozyme sensor for NO was explored using CV (Figure 3c and S8, Supporting Information). The CV current response gradually increased with decreasing acidity and reached a maximum at pH 7.4, and decreased when the electrolyte favored alkaline. Results indicated Ce$_1$-CN behaved the highest catalytic activity and interaction with NO molecules at pH 7.4, which was chosen for subsequent studies. As shown in Figure 3d, the applied voltage was regulated using the amperometric I-t method, and it was found that the electrochemical sensitivity reached the highest value with the gradual increase of the applied voltage from 0.8 V to 0.95 V. When the applied voltage was further increased to 1.0 V, the current response showed a decreasing trend. Therefore, the applied voltage of 0.95 V was selected as the optimum potential.

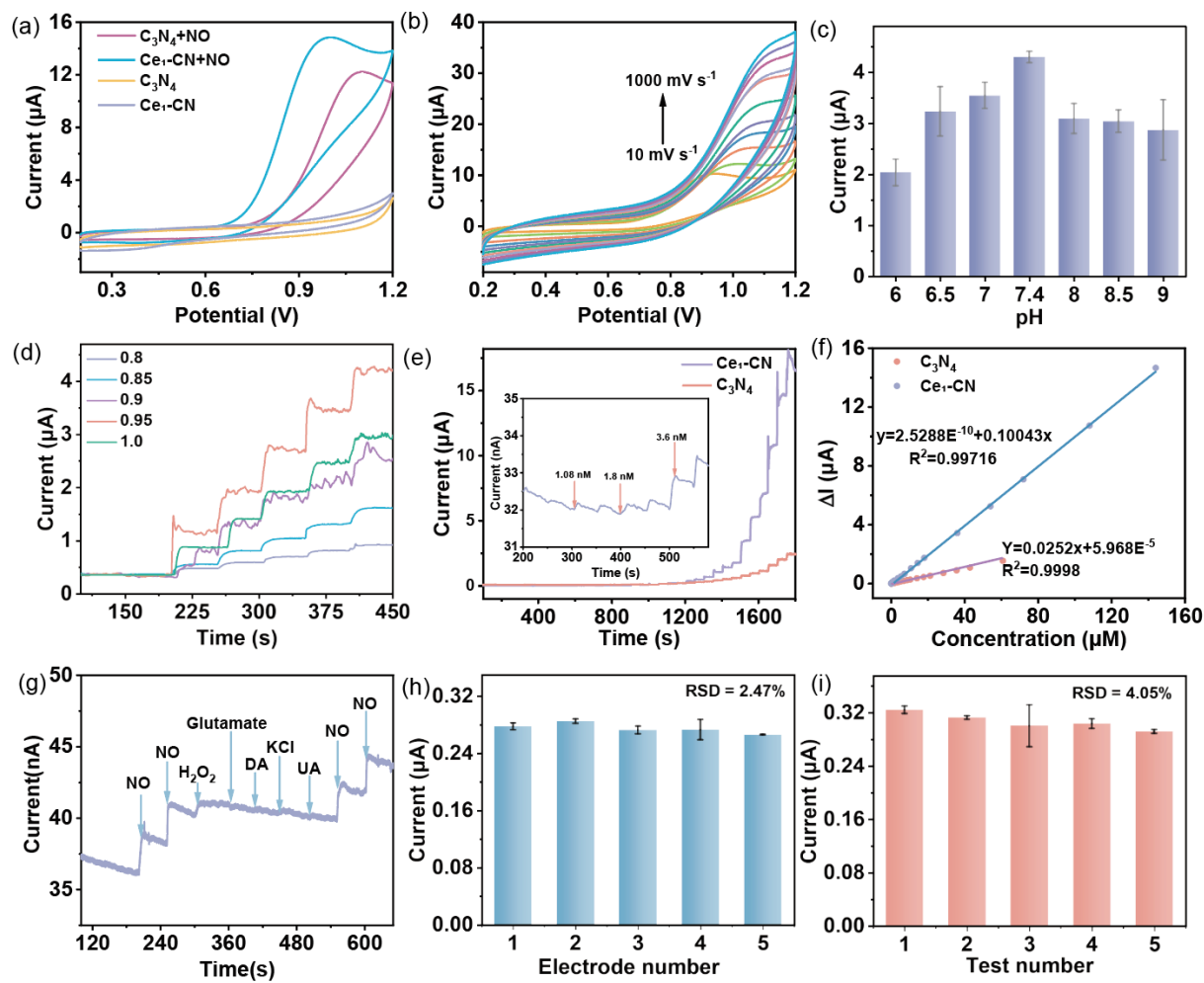

**Figure 3.** (a) CV response of $C_3N_4$ and $Ce_1$-CN toward 72 μM NO in 0.01 M PBS. (b) CV response curves of $Ce_1$-CN sensor to 36 μM NO at different scan rates varing from 10 to 1000 mVs$^{-1}$. Influence of (c) pH and (d) applied potential on current response of $Ce_1$-CN toward NO. (e) Amperometric response of the $Ce_1$-CN and $C_3N_4$ biosensor to successive addition of different concentrations of NO at an applied potential of 0.95 V. (f) Linear calibration curve of the $Ce_1$-CN and $C_3N_4$ biosensor toward NO sensing. (g) Interference testing of $Ce_1$-CN biosensor. (h) Reproducibility and (i) Repeatability of $Ce_1$-CN biosensor.

The electrochemical properties of $Ce_1$-CN were further investigated by the amperometric I–t method, recorded at a potential of 0.95 V to study the response of continuously added NO in 0.01 M PBS with pH 7.4. The I–t curve shows that the sensor quickly generates a step response after addition of NO (purple line Figure 3e). The amperometric current measured at 0.95 V was linearly and positively correlated with the increase in NO concentration. The linear equation for $Ce_1$-CN was $\Delta I$ (A) = 2.5288×10$^{-10}$ + 0.10043$C_{NO}$ (M) (R = 0.997,), with an ultrawide linear concentration range of 1.08 nM to 143.9 μM, and the detection limit (LOD) is 0.36 nM (blue line Figure 3f). It is obtained from the expression LOD = 3 S/K, where S is the standard deviation of the blank signal and K is the sensitivity estimated from the slope of the calibration curve. Compared with the performance of $C_3N_4$ (red line Figure 3f), which obtained a linear range from 18 nM to 60.82 μM with a LOD of 6 nM (Figure S9, Supporting Information), $Ce_1$-CN biosensor performed a two orders of magnitude wider linear range and a 16.7 times lower LOD. Results indicate that the electrochemical response of $Ce_1$-CN is significantly higher than that of $C_3N_4$. Besides, the performance of the $Ce_1$-CN biosensor is much better than the reported previous works (Table S1, Supporting Information). The sensitivity of $Ce_1$-CN sensor was calculated as 1434.7 nA μM$^{-1}$ cm$^{-2}$. The response time of $Ce_1$-CN sensor for NO is 1.7 s (Figure S10, Supporting Information), which can quickly reach the steady state current. Due to the half-life time of NO molecule is about 3-6 s, therefore, the sensor can quickly capture the signal of NO, and has broad application prospects for detecting NO in vitro.

To further evaluate the activity of $Ce_1$-CN nanoenzyme, Michaelis-Menten constant (Km) is investigated. The value of the Km is equal to the concentration of substrate at which the enzymatic reaction reaches half of its maximum velocity, which means that the larger the Km, the more substrate is needed to reach the maximum velocity. So the smaller the Km, the better the enzyme catalyzes[36]. Based on the Lineweaver-Burk plot, the Km of $Ce_1$-CN was calculated to be 2.07 nM (Figure S11, Supporting Information). Whereas the reported Km of the NO sensor fabricated by iron-based mimetic enzyme, carboxylated single-walled carbon nanotubes and $CeO_2$ were 4610 nM [37], 4300 nM [38] and 10.88 nM [39], indicating $Ce_1$-CN behaved high biomimetic catalytic activity. In biological environments, the detection of NO is usually interfered by other substances, such as $H_2O_2$, DA, UA, so it is necessary to investigate the effect of these interfering substances on the sensor. The selectivity of the $Ce_1$-CN sensor can be realized by detecting the current response of various interfering species that may coexist with NO. Results showed that the weak current response induced by adding 18 mM of $H_2O_2$, glutamate, $K^+$, $Cl^-$, UA and physiological concentrations of DA ($10^{-8}$ M) to the system did not interfere with the detection of 1.8 mM NO (Figure 3g). It can be concluded that the sensor has good anti-interference ability, and can adapt to the biological environment. The reproducibility (Figure 3h) and repeatability (Figure 3i) of $Ce_1$-CN biosensor were estimated through amperometric I-t, the relative standard deviation (RSD) was calculated to be 2.47 % and 4.05 %, respectively, indicating good performance. Besides, the proposed $Ce_1$-CN biosensor also behaved satisfactory stability toward NO sensing, which can maintain 87.2 % of the initial response after 14 days (Figure S12, Supporting Information).

## 2.3. Catalytic reaction mechanism of $Ce_1$-CN/GCE toward NO

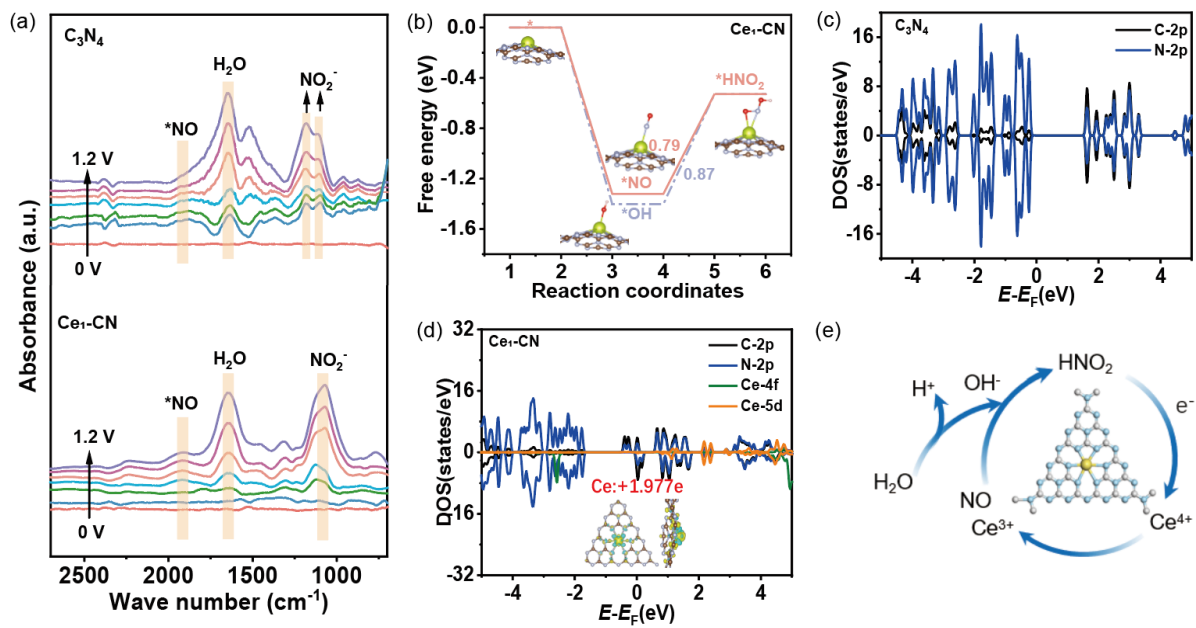

**Figure 4.** (a) ATR - SEIRAS spectra recorded NO adsorption and reaction on $Ce_1$-CN and $C_3N_4$ through the addition of NO to a nitrogen saturated 0.01 M PBS solution. (b) Schematic Gibbs free energy diagrams of the $Ce_1$-CN and the adsorption configurations of *NO, *OH, and *$HNO_2$. Density of States of (c) $C_3N_4$ and (d) $Ce_1$-CN. Inset of (d) Differential charge density of $Ce_1$-CN. The yellow and blue area represented charge loss and charge accumulation, respectively, the iso-surface is 0.01 e/$A^3$. (e) Diagram of the reaction mechanism.

To eliminated the oxidation process of NO on the surface of electrode materials, in situ attenuated total reflection surface-enhanced infrared spectroscopy (ATR-SEIRAS)[40] was utilized to identify the intermediate species formed during electrochemical oxidation of NO over catalyst. By comparing with the ATR-SEIRAS data obtained of NO over $Ce_1$-CN and $C_3N_4$ under a series of applied potentials ranging from 0 V to 1.2 V (vs. Ag/AgCl) (**Figure 4a** and S13, Supporting Information), vibration peaks associated with reaction intermediates were observed at 1930 $cm^{-1}$ and approximately 1100 $cm^{-1}$, corresponding to adsorbed NO (*NO) and nitrite species ($NO_2^-$), respectively[41-42]. Comparative analysis of the ATR-SEIRAS spectra of $Ce_1$-CN and $C_3N_4$ revealed that $Ce_1$-CN exhibits a significantly enhanced *NO adsorption capacity relative to $C_3N_4$, as indicated by the progressive increase in peak intensity observed during the reaction process. These findings further support the conclusion that $Ce_1$-CN demonstrates superior adsorption and activation performance toward NO. In addition, a red shift in the $NO_2^-$ signal of $Ce_1$-CN relative to $C_3N_4$ was observed, which may be attributed to weaker adsorption of the $NO_2^-$ bond on $Ce_1$-CN, potentially facilitating a faster subsequent reaction. Furthermore, the intermediate species during oxidation of NO was also confirmed by an ion concentration detection method with p-aminobenzenesulfonamide as color reagent using UV-Vis spectrophotometry (Figure S14, Supporting Information). Results showed the color of p-aminobenzenesulfonamide changes from transparent to deep red after incubation with the electrochemical reaction mixture of $Ce_1$-CN and NO, indicating the production of $NO_2^-$, further confirming the results of ATR-SEIRAS.

Density functional theory (DFT) calculations were further conducted to understand the NO oxidation reaction mechanism over electrode materials. Two possible reaction pathways, as shown in Figure 4b and S15, were considered. Path one is the NO first adsorbs on the catalyst, and then reacts with hydroxyl anion ($OH^-$), and path two is the reverse process. Results showed that both NO and $OH^-$ species are very hard to adsorb on $C_3N_4$, requiring high adsorption free energies as 0.92 eV and 1.97 eV, respectively. After loading Ce atom into $Ce_1$-CN, the adsorption Gibbs free energies for NO and $OH^-$ species significantly decreased to -1.32 eV and -1.40 eV, indicating the reaction is more favorable over $Ce_1$-CN compared with $C_3N_4$. For the subsequent reaction between *NO and *$OH^-$ over $Ce_1$-CN, 0.79 eV and 0.87 eV free energies are required for path one and two, respectively. Thus, path one is energetically favorable for $Ce_1$-CN. Results showed that the rate-determining step Gibbs free energies for $C_3N_4$ and $Ce_1$-CN system were 0.92 eV and 0.79 eV, indicating $Ce_1$-CN behaved higher activity than $C_3N_4$. The density of states (DOS) analysis of samples demonstrated that $C_3N_4$ (Figure 4c) exhibits semiconductor characteristics with no electronic state at the Fermi level, leading to poor activity. However, for $Ce_1$-CN (Figure 4d), there is an electronic spike at the Fermi level including contributions from C 2p, N 2p and Ce 5d orbitals, indicating the Ce 5d might be the activity orbital. The presence of Ce 5d and N 2p inner-layer electronic states suggests an electronic coupling between Ce and N. Besides, the differential charge density analysis (inset of Figure 4d) indicated that Ce atom of $Ce_1$-CN lost 1.977e after coordination with N atom, which is benefit for the oxidation of NO. Based on in-situ ATR-SEIRAS measurements and DFT calculations, the excellent NO detection performance of $Ce_1$-CN nanozyme can be attributed to the strong coupling of d-f electron orbitals, which ensures a rapid cycle between $Ce^{3+}$ and $Ce^{4+}$. The mechanism of NO electrochemical oxidation over $Ce_1$-CN was shown in Figure 4e.

**2.4. Evaluation of in vitro NO sensing generated by trigeminal nerve of rat with Ce$_1$-CN biosensor**

The trigeminal nerve arises from the trigeminal ganglion and divides into three major branches: the ophthalmic nerve (V1), maxillary nerve (V2), and mandibular nerve (V3), listed in rostral-to-caudal sequence. It has been shown that nitric oxide synthase (NOS) plays a key role in the trigeminal pain signaling pathway[43-44]. This is owing to arginase 2 (ARG2) and NOS competitively utilize their shared substrate, L-arginine, thereby reciprocally modulating NO biosynthesis[45] (**Figure 5a**). Overproduction of NO will cause TN. Carbamazepine is the first-line therapeutic agent for trigeminal neuralgia, which can stabilize the overexcited nerve cell membrane, reduce the conduction of nerve impulses, and inhibit the transmission of pain signals in the trigeminal nerve[46-47]. In this work, we established three experimental groups: (i) sham-operated controls (undergoing surgical simulation without neural injury), (ii) TN model rats exhibiting characteristic neuropathic pain behaviors and pathophysiological alterations, and (iii) TN model rats receiving carbamazepine treatment (5mg/kg). The trigeminal nerve of the rat was divided into three branches during sensing (Figure S16, Supporting Information).

To validate whether the TN rat model was successful, mechanical withdrawal threshold testing (Figure 5b-d) and face-grooming frequency behavior (Figure 5e and Video S1, Supporting Information) were conducted. The calculated p-values for V1, V2, and V3 were $2.27 \times 10^{-5}$, $1.10 \times 10^{-4}$, and $2.15 \times 10^{-4}$, respectively. Since all were significantly less than 0.05, this indicates that the mean of each dataset (V1, V2, and V3) significantly differs from zero. Thus, the TN rat model was successfully built. The pain threshold was expressed as median (interquartile range, IQR) in the form of a (b-c), where a represents the median of all data, and b and c correspond to the median of the first 50% of the data (lower quartile, Q1) and the median of the latter 50% of the data (upper quartile, Q3), respectively. Results showed the pain thresholds of sham operation rats (normal rats) were 2.00 g (1.40-8.00), 1.00 g (1.00-2.00) and 1.40 g (1.00-2.00) at the trigeminal nerve of sham-V1, sham-V2 and sham-V3, respectively, 5 days after operation. The pain thresholds of TN-V1, TN-V2 and TN-V3 in trigeminal nerve of TN rats were 0.60 g (0.4-1.00), 0.40 g (0.16-0.40) and 0.16 g (0.07-0.40) respectively. This indicates that the pain thresholds of the trigeminal nerve in TN rats are significantly lower than those in normal rats. Besides, it can be seen that on the first day after carbamazepine treatment, the pain thresholds in the three parts of trigeminal nerve of TN rats were significantly increased from 0.04 g (0.008-0.08), 0.04 g (0.02-0.04), 0.04 g (0.04-0.07) before administration to 1.40 g (0.60-2.00), 1.40 g (1.00-2.00), 1.40 g (1.40-2.00). This showed that the symptoms of trigeminal neuralgia were alleviated after treatment.

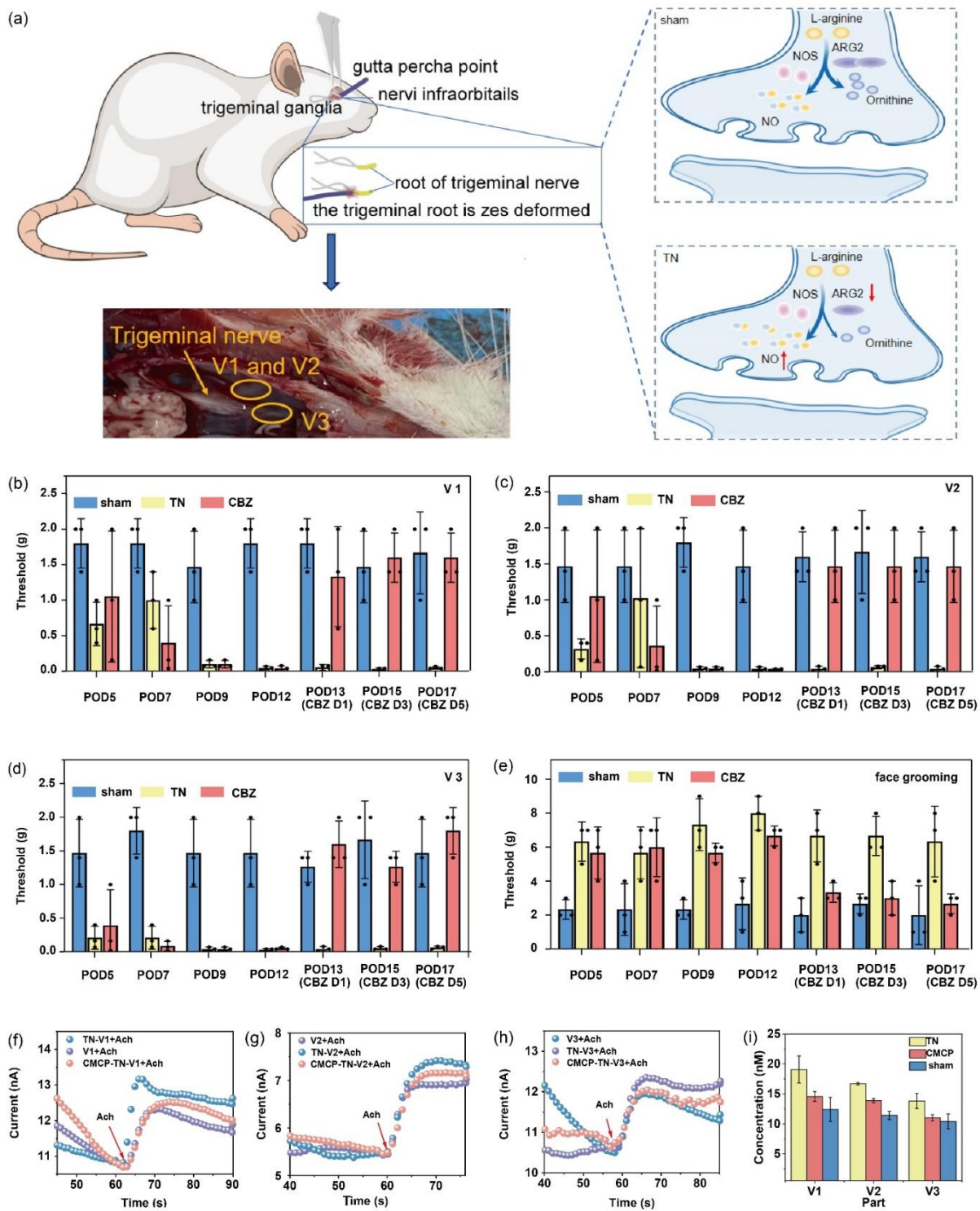

**Figure 5.** (a) Schematic illustration of the role of NO in TN. Changes in mechanical withdrawal threshold in different rat models (b) V1; (c) V2; (d) V3. (e) Changes in mechanical withdrawal threshold in TN rat model after carbamazepine treatment. Amperometric I-t curves of $Ce_1$-CN/GCE to monitor the NO produced by different part of the trigeminal nerve in rats:(f) V1; (g) V2; (h) V3. (i) The column diagram of the concentration of NO generated by three branches of trigeminal nerve in TN rats, carbamazepine treated TN rats and normal rats.

We further investigated the NO production from three parts of the trigeminal nerve in normal, TN and carbamazepine treated rats (Figure 5f-h), respectively. For amperometric I-t curves, as soon as 0.5 mM Ach (a drug commonly used to stimulate cells to generated NO) was injected into the 0.01 M PBS pH 7.4 containing different trigeminal nerve branches, distinct oxidizing currents step were observed due to the electrochemical reaction of the NO molecules generated by different trigeminal nerve branches (Figure S17-22, Supporting Information). The corresponding NO concentrations were calculated according to the current responses (Figure S23, Table S2, Supporting Information) and linear equation obtained by Figure 3f blue line, which were summarized in Figure 5i and Table S3. Results indicated that compared with normal rats, the current responses and NO concentrations in each branch (V1,V2 and V3) of the trigeminal nerve in TN rats were significantly increased. After carbamazepine treatment, both parameters decreased compared with the TN group and approached to normal levels. It can be seen that the concentration of NO produced by trigeminal nerve V1, V2 and V3 in TN rats is 1.41, 1.46 and 1.28 times of that produced by normal rats, respectively. This may be attributed to the low regulation of ARG2 expression at the onset of trigeminal neuralgia, which allows NOS to access more L-arginine and thereby synthesize higher levels of NO[48]. After carbamazepine treatment, the concentration of NO produced by trigeminal nerve V1, V2 and V3 in rats was 1.17, 1.22 and 1.02 times that produced by normal trigeminal nerve V1, V2 and V3, respectively. Thus, carbamazepine treatment can reduce the release of NO from the trigeminal nerve in TN. Combined with the threshold testing and face-grooming behavior results, it can be concluded carbamazepine treatment can reduce the pain of rats, and the symptoms were relieved, which proves the role of NO in trigeminal neuralgia.

## 3. Conclusion

In summary, we developed an electrochemical sensor based on Ce single atom nanoenzyme for the detection of NO. Electrochemical analyses demonstrated that the sensor exhibited high performance, including rapid response kinetics, achieving a steady-state current within 1.7 s. Utilizing this sensor, we successfully quantified NO release from different parts of the trigeminal nerve in rats with TN after stimulation with Ach. The results indicated that the NO release from the trigeminal nerve of TN rats was higher than that in normal rats, suggesting that abnormal fluctuation in NO level may be closely related to the TN. After the treatment with carbamazepine, a notable decrease in NO release was observed, accompanied by a reduction in pain-related behaviors in TN rats. The proposed NO biosensor provides a promising tool for investigating complex biological processes and underlying disease mechanisms.

**Acknowledgements**

This work was supported by the National Natural Science Foundation of China (22475145, 21705115, 82401149), and the Natural Science Research Foundation of Jiangsu Higher Education Institutions (23KJB150034). The Qinglan Project of Jiangsu Province for F.X.H. and the Jiangsu Government Scholarship for Overseas Studies. L.G.O. acknowledges funding from the Engineering and Physical Sciences Research 446 Council (EPSRC) grants No. EP/K03099X/1, EP/W024284/1.

**Data Availability Statement**

The data used in the current study are available from the corresponding authors on reasonable request.